\documentclass[10pt,notitlepage,aps,prb,reprint,superscriptaddress,amsmath,amssymb,floatfix,showpacs]{revtex4-1}

\usepackage{float}
\usepackage[utf8]{inputenc}
\usepackage{graphicx}
\usepackage{dcolumn}
\usepackage{bm}
\usepackage[english]{babel}
\usepackage{textcomp}
\usepackage[font=scriptsize]{caption}
\usepackage{amsmath}
\usepackage{color,soul}
\begin{document}
\preprint{APS/123-QED}
\title{Searching for the signature of a pair density wave in YBa$_2$Cu$_3$O$_{6.67}$ using high energy X-ray diffraction}

\author{E. Blackburn}
\email{elizabeth.blackburn@sljus.lu.se}
\affiliation{Division of Synchrotron Radiation Research, Lund University, SE-22100 Lund, Sweden}
\author{O. Ivashko}
\affiliation{Physik-Institut, Universit\"at Z\"urich, Winterthurerstrasse 190, CH-8057 Z\"urich, Switzerland}
\author{E. Campillo}
\affiliation{Division of Synchrotron Radiation Research, Lund University, SE-22100 Lund, Sweden}
\author{M. v. Zimmermann}
\affiliation{Deutsches Elektronen-Synchrotron DESY, 22607 Hamburg, Germany}
\author{R. Liang}
\affiliation{Department of Physics \& Astronomy, University of British Columbia, Vancouver, Canada}
\affiliation{Canadian Institute for Advanced Research, Toronto, Canada}
\author{D. A. Bonn}
\affiliation{Department of Physics \& Astronomy, University of British Columbia, Vancouver, Canada}
\affiliation{Canadian Institute for Advanced Research, Toronto, Canada}
\author{W. N. Hardy}
\affiliation{Department of Physics \& Astronomy, University of British Columbia, Vancouver, Canada}
\affiliation{Canadian Institute for Advanced Research, Toronto, Canada}
\author{J. Chang}
\affiliation{Physik-Institut, Universit\"at Z\"urich, Winterthurerstrasse 190, CH-8057 Z\"urich, Switzerland}
\author{E. M. Forgan}
\affiliation{School of Physics and Astronomy, University of Birmingham, Birmingham, B15 2TT, United Kingdom}
\author{S. M. Hayden}
\email{s.hayden@bristol.ac.uk}
\affiliation{H.~H.~Wills Physics Laboratory, University of Bristol, Bristol, BS8 1TL, United Kingdom}

\date{\today}

\begin{abstract}
We have carried out a search for a pair density wave signature using high-energy X-ray diffraction in fields up to 16 T.  We do not see evidence for a signal at the predicted wavevector.  This is a report on the details of our experiment, with information on where in reciprocal space we looked.
\end{abstract}

\maketitle

\section{Introduction}

In a scanning tunnelling microscopy experiment, Edkins {\it et al.} \cite{Edkins2019_EKFM} reported the observation of two periodic electron-density waves in the haloes around the vortex cores in the cuprate superconductor Bi$_2$Sr$_2$CaCu$_2$O$_8$.  This is observed in differential tunnelling conductance maps of the sample for energies between 25 and 45 meV.  One of these electron-density waves corresponds to the charge density wave (CDW) previously reported in this material and many other cuprates 
Based on the relationship between the two electron-density waves, the second signal is inferred to be a secondary signature from an underlying Cooper-pair density wave (PDW) state that induces the previously observed charge density wave state.

The Cooper-pair density wave state has been a subject of increasing interest over the years.  The most well-known example of such a state is the Fulde-Ferrell-Larkin-Ovchinnikov (FFLO) state\cite{LO,FF} predicted to occur in a magnetic field, and inferred to exist in certain organic superconductors 
In this state, the degeneracy of the spin-up and spin-down Fermi surfaces is broken, giving rise to Cooper pairs with a finite momentum. The PDW state is much more general than this, and does not require breaking of time reversal symmetry, instead merely requiring that the superconducting order parameter (the gap) varies periodically in space, with a spatial average of zero.\cite{Agterberg2019}  If a given PDW state, with a given set of symmetries, may exist, it will exist, although the question of the energy scales remains unknown.  The specific symmetries of the state will generate different types of induced order.  

The phase diagrams of the cuprate superconductors are rich in density wave states of many different types.  The CuO$_2$ planes that are the seat of superconductivity in these materials are very sensitive to their local environment, manifesting many different types of instability.  A number of these instabilities may be described using a `density wave' description, where a given parameter/degree of freedom (e.g.~charge, spin) varies periodically over a given lengthscale.  The most well-known of these instabilities are perhaps the charge and spin stripes seen in (La,Ba)$_2$CuO$_4$.\cite{Tranquada2013} 

The Cooper-pair density wave state constitutes one proposal to unify this picture.  The PDW state is posited as the `parent' phase, which breaks multiple symmetries. The various experimental observations of other density-wave states are then `daughter' phases, corresponding to the partial melting of the parent phase, such that the observed phenomena correspond to a subset of the symmetries of the parent phase.\cite{Fradkin2015_FrKT,Agterberg2019}

This proposal is, however, hard to verify, as it is experimentally challenging to identify the direct fingerprint of the pair density wave state.  Several potential tests have been identified \cite{Berg2009_BeFK,Fradkin2015_FrKT}; the approach taken by Edkins {\it et al.} was to look for a charge density modulation that is generated by the underlying pair density wave.  Scanning Josephson tunnelling microscopy has now been used successfully in, for example, NbSe$_2$, where the CDW and PDW are observed at the same wavevector.\cite{Liu}

\section{The relationship between the CDW and the PDW}

The pair density wave state is defined as a spatial modulation in the superconducting gap function.  This will naturally give rise to a variation in the spatial electron density, $\propto | \Delta_{PDW}(\bf{r}) |^2$, with a wavevector twice that of the underlying PDW wavevector, $\bm{q}_{CDW} = 2 \bm{q}_{PDW}$.  Such an induced charge density wave is detectable by multiple methods, including scanning tunnelling microscopy (STM), and X-ray diffraction.  However, from the observation alone, it is impossible to determine whether a given CDW signal comes from an independent CDW order parameter, or is induced by an underlying PDW state.

Accordingly, in the cuprate superconductors there must also be a uniform superconducting order parameter, $\Delta_0$, which by observation, is primarily of $d$-symmetry.  This introduces a cross-term into the spatial electron density, giving rise to an induced charge density wave with the same wavevector as the underlying pair density wave state, $\bm{q}_{PDW}$, as can be seen by considering the effect of a model superconducting gap function $\Delta_0 + \Delta_{PDW}\sin{\bf{q} \cdot \bf{r}}$ on the electron density:

\begin{eqnarray}
    \rho_e(r) \propto |\Delta_0|^2 & + 2 \Delta_0 \Delta_{\text{PDW}} \sin(\bm{q}_{\text{PDW}} \cdot \bm{r}) \nonumber \\
   & + | \Delta_{\text{PDW}} |^2 \sin^2(\bm{q}_{\text{PDW}} \cdot \bm{r}).
\end{eqnarray}
This is obviously an oversimplified model; for a more detailed treatment, there is an excellent review by Agterberg {\it et al.}\cite{Agterberg2019}  The key aspect to take from this is the relationship between the observed wavevectors: $\bm{q}_{CDW} = 2 \bm{q}_{PDW}$.  This is the primary observation in the work by Edkins {\it et al.}\cite{Edkins2019_EKFM}  

\begin{figure}
	\centering
	\includegraphics[width=0.4\textwidth]{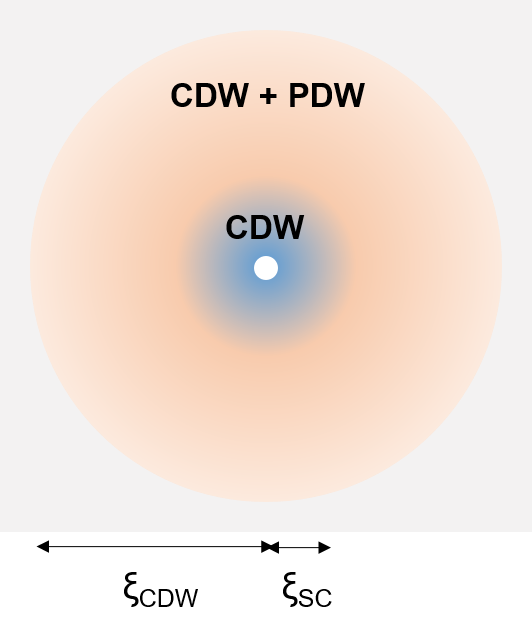}
	\caption{Sketch of the proposed vortex halo based on the proposal of Agterberg and Garaud.\cite{AgterbergGaraud} 
 }
	\label{fig:vortexhalo}
\end{figure}

Edkins {\it et al.} report that the PDW signature is confined to areas around the vortex cores.  
To understand why, we must consider the relative strength of the various order parameters.  The ordinary superconducting order parameter, $\Delta_0$, will be completely suppressed at the vortex core, developing its usual value on a lengthscale determined by the penetration depth.  Various authors suggest that $\Delta_{PDW}$ will appear where $\Delta_0$ is suppressed, although this is not necessary.\cite{Wang2018_WEHD,Edkins2019_EKFM,Agterberg2019}  In Figure \ref{fig:vortexhalo} we illustrate the consequences of these variations on the observed CDW and PDW signals.  A secondary consequence of these effects is that the temperature dependence of the two signals should be different, with the signal at ${\bf q}_{PDW}$ disappearing above $T_C$, whereas the signal at ${\bf q}_{PDW}$ may persist to a higher temperature.

Here we present details of our investigation of a sample of YBa$_2$Cu$_3$O$_{6.67}$ (YBCO) using X-ray diffraction to look for the induced pair density wave signal at ${\bf q}_{PDW}$.  This is a different experimental technique to scanning tunnelling microscopy, and so we will discuss briefly what is actually probed in the X-ray diffraction experiment.
We will then describe the experimental results.

\section{What do X-rays actually see?}

In our experiment, we use high energy (98.2 keV) X-rays.  The X-rays scatter off the overall charge distribution of the electrons in the sample, with contributions from each atom.  The contributions from each element have a particular form factor, $f(Q)$, determined by the electronic orbital shape and size distribution.  For a classical charge density wave, we may consider that there is a spatial variation of charges, independent of the lattice.  To describe this from an atomistic starting point, we can consider that the charge associated with a given atom (this can also be thought of as the valence of that atom) may vary in space.  In addition to this charge distribution, there will also be a corresponding shift in atomic positions, through electron-phonon coupling.  Of course, in a given material, the atomic displacements may occur first, and be followed by a charge distribution.

High energy X-ray diffraction is most sensitive to atomic displacements, as the X-rays will scatter primarily off the large numbers of core electrons associated with the individual atoms, swamping any signature from small charge variations.\cite{Chang2012_CBHC}  

In the picture described in Figure \ref{fig:vortexhalo}, any signal to be observed at ${\bf q}_{PDW}$ will be confined to regions with a correlation length determined by the size of the vortex halo, following the observations of Edkins \textit{et al.}.\cite{Edkins2019_EKFM}  We will therefore only be able to see this signal at high fields (as is also the case for the STM measurements).  If we take this correlation length as being $\sim 100$ \AA, we will be able to see the effects of this on the width of the resulting Bragg peak, as it will result in a finite width determined by this correlation length.  Edkins {\it et al.} have argued that this width should be half that of the signal at ${\bf q}_{CDW}$ (see Figure \ref{fig:vortexhalo}).\cite{Edkins2019_EKFM}  We should expect to see Bragg peaks due to ${\bf q}_{PDW}$ for a large range of momentum transfer values, as this is determined by the underlying atomic form factors.  
\begin{figure}[t]
	\centering
    \includegraphics[width=0.49\textwidth]{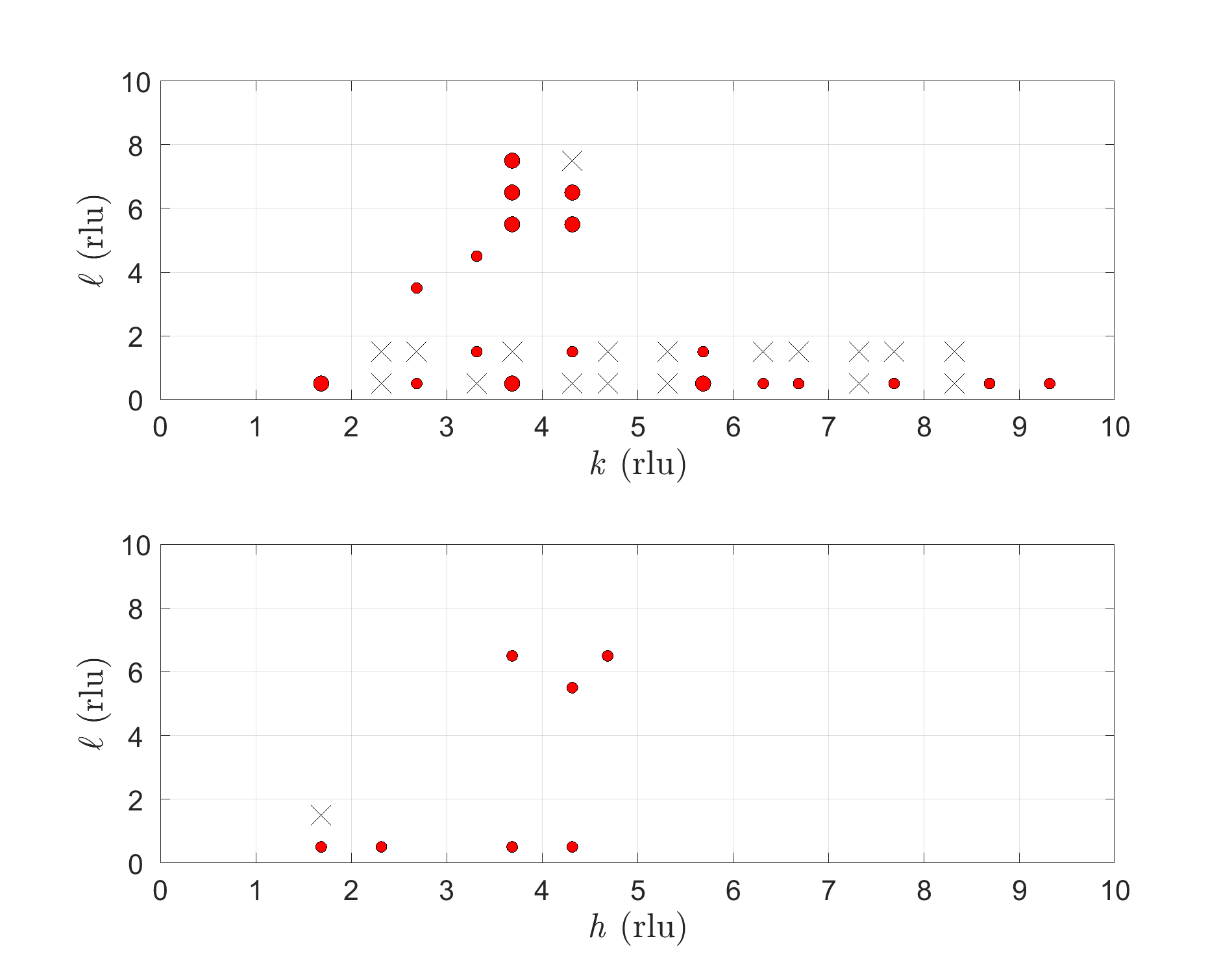}
	\caption{Positions in reciprocal space studied during this experiment, in the (upper) $(0KL)$ plane and (lower) $(H0L)$ plane.  A cross indicates that there was no clear signal at ${\bf q}_{CDW}$ at 0 T.  The large red circles indicate that there was a clear signal at ${\bf q}_{CDW}$ , and the small red circles indicate either a small signal, or the appearance of unexpected background across the scan.  The axes are drawn to scale.}
	\label{fig:dotplot}
\end{figure}

\section{Experimental Methods}

We have carried out a high energy X-ray diffraction experiment to look for signs of the CDWs induced by a potential PDW in YBCO.  These hard X-ray (98.2 keV) diffraction experiments were carried out at the P07 triple-axis diffractometer at the PETRA-III synchrotron at DESY (Hamburg, Germany).  The sample of YBa$_2$Cu$_3$O$_{6.67}$, with an ortho-VIII structure, has previously been characterized and used in Refs.~\onlinecite{Chang2012_CBHC,Chang2016_CBIH}).  The sample was mounted in a horizontal 17 T cryomagnet.\cite{Holmes2012_HWBF}  The setup was identical to that used in Ref.~\onlinecite{Chang2016_CBIH}.  The sample was mounted to access either the $(0,K,L)$ or $(H,0,L)$ scattering planes.  A second sample with an ortho-II structure was briefly studied, but not in as much depth, and so we do not present any results here.

Figure \ref{fig:dotplot} indicates the regions of reciprocal space studied during the experiments, in the (0 K L) and (H 0 L) planes.  Although the figure only denotes points at half-integral $L$ positions, corresponding to the previously observed CDW wavevectors, measurements were also made at associated integral $L$ positions, as well as at a selection of other $L$ values.  Access to reciprocal space is limited by two conditions.  Firstly, if the sample is aligned such that field is parallel to the $c$ axis, the opening angles of the cryomagnet restrict the maximum $2\theta$ to 20$^{\circ}$.  This condition applies to the points measured for $L < 2.5$.  Secondly, the sample may be rotated inside the cryomagnet, such that the field is no longer parallel to the $c$ axis.  This was done to access the points at higher $L$ values.  This has the disadvantage of limiting the field along the $c$ axis.  We measured all of the points accessible with the field along the $c$ axis in the (0 K L) plane, and with the sample rotated with respect to the field, we selected points in a region where our previous experiments\cite{Forgan2015_FBHB} had indicated that the CDW signal was strong.

\begin{figure}[t]
	\centering
	\includegraphics[width=0.49\textwidth]{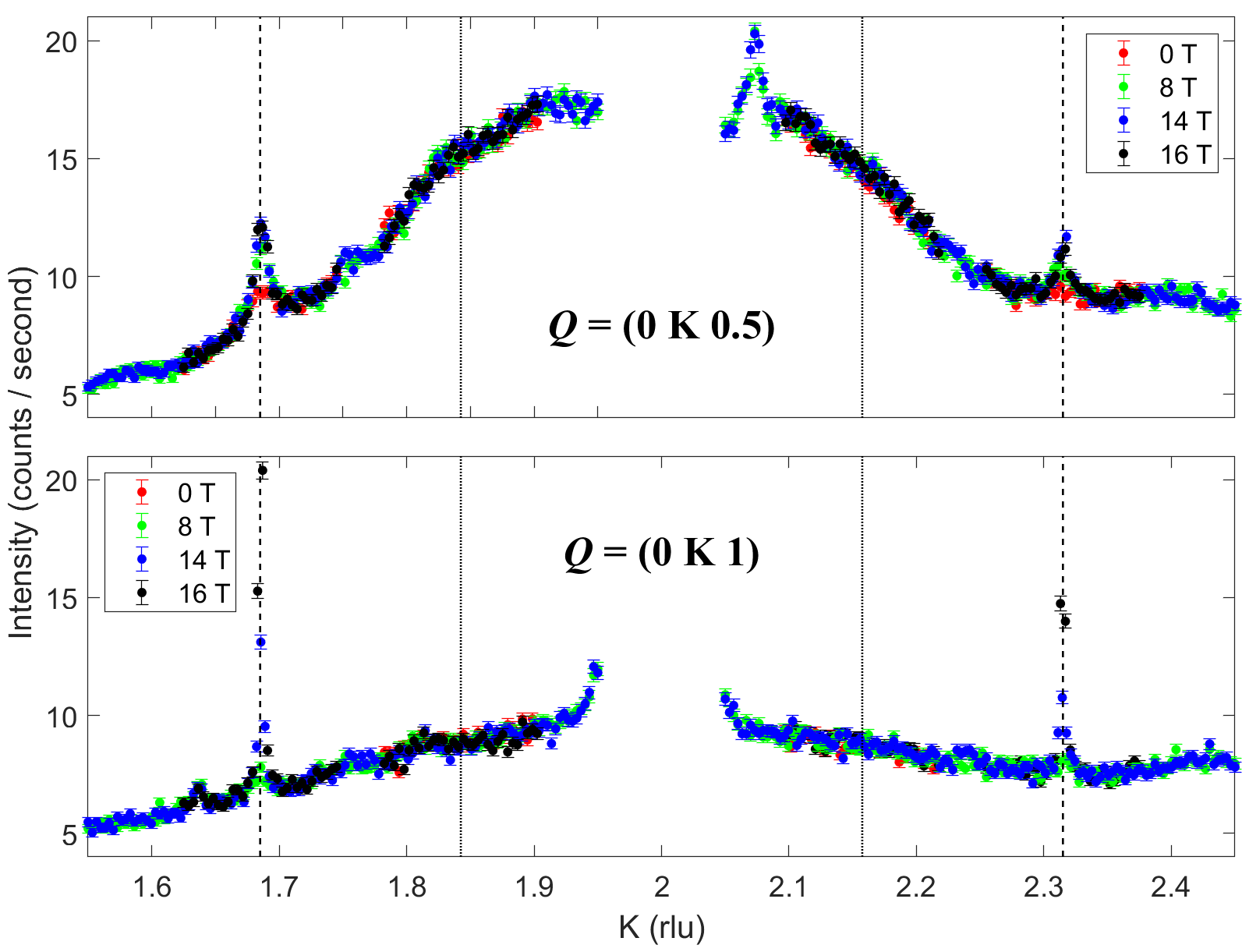}
	\caption{Scattered intensity at (0 $K$ $L$) in fields from 0 T to 16 T, applied parallel to the $c$ axis.  The upper panel shows $L$ = 0.5 and the lower panel $L$ = 1.  The dashed lines indicate the observed $q_{CDW}$ and the dotted lines indicate the predicted $q_{PDW} = \frac{1}{2}q_{CDW}$.}
	\label{fig:KLcuts}
\end{figure}

At certain positions in reciprocal space, our measurements show peaks that come from structured diffuse scattering, as reported by Le Tacon \textit{et al.};\cite{LeTacon2013_LBSD} a trace of this can be seen in the upper panel of Figure \ref{fig:KLcuts}.  This scattering is identified as such by scanning towards the parent Bragg peak and observing a continual increase.

\begin{figure}
	\centering
	\includegraphics[width=0.49\textwidth]{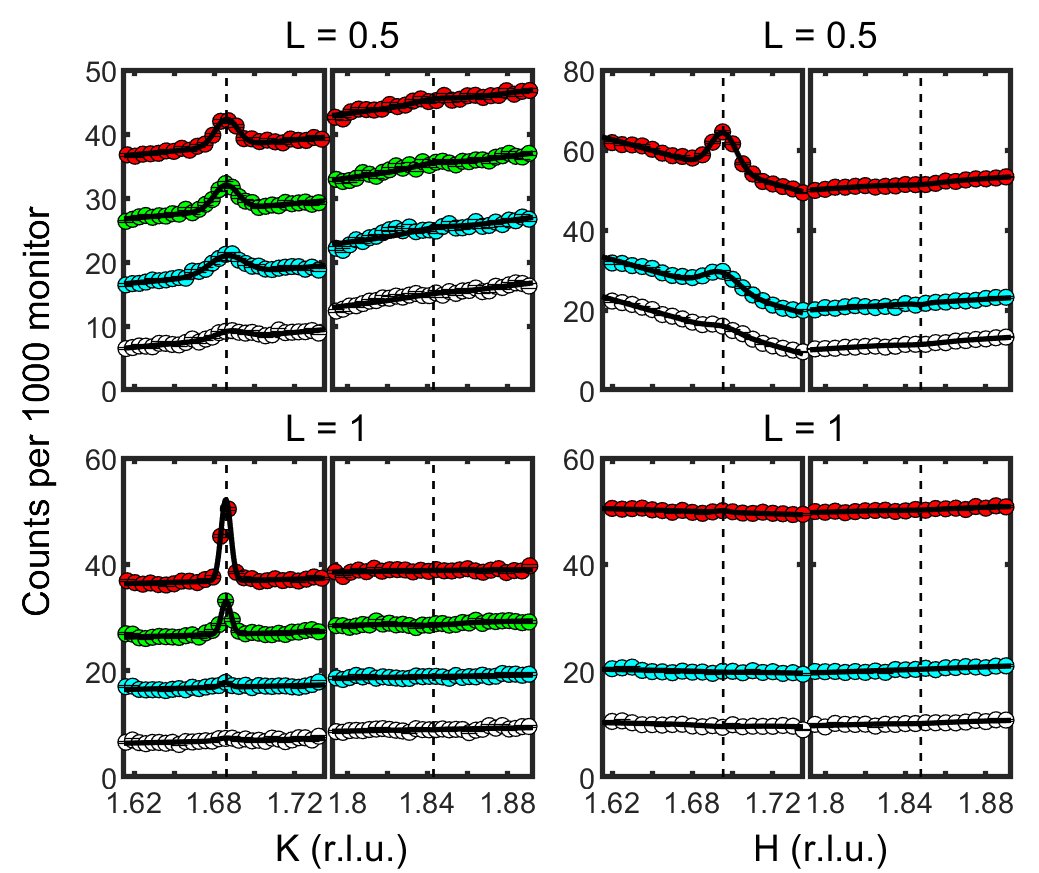}
	\caption{Cuts through (left) $K$ and (right) $H$ at the CDW and PDW positions.  The upper panels are measured at $L$ = 0.5, and the lower panels at $L$ = 1.	A monitor of 1000 corresponds to a counting time of approximately 1 second.  The dashed lines denote the CDW and PDW positions respectively.  Several fields are shown in each panel: white circles are 0 T, blue circles are 8 T, green circles are 14 T, and red circles are 16 T.  The curves at different fields are separated by 10 counts up the vertical axis.}
	\label{fig:linecuts_all}
\end{figure}

\section{Results}

We show here data from a subset of the regions studied in Fig.~\ref{fig:dotplot}, choosing to focus on the data collected close to either $H$ or $K$ = 2 and $L \le 1$.  These positions give us the lowest overall magnitude of momentum transfer $Q$, and correspond to a strong CDW peak.  The results presented here are characteristic of the results obtained at the other positions studied.

Figure \ref{fig:KLcuts} gives a broad overview of the data obtained in the $(0KL)$ plane around the (0 2 0.5) and (0 2 1) positions, in fields from 0 T to 16 T.  The charge density wave is observed in its usual position ($q_{CDW}$ = 0.315 \AA $^{-1}$).  The charge density wave reflection is visible at 0 T only at (0 2-$q_{CDW}$ 0.5), but application of field makes it visible at (0 2+$q_{CDW}$ 0.5), and we can also see the field-induced reflections at $L$ = 1 as well.  We do not see any sign of increased scattering at the position $q_{PDW} = \frac{1}{2}q_{CDW}$.  The peak seen at (0 2.05 0.5) comes from the structured diffuse scattering noted above.  The overall shape of the background scattering is similar to that observed in our previous experiments on multiple samples, see e.g.~\onlinecite{Forgan2015_FBHB}. 

\begin{figure*}
	\centering
	\includegraphics[width=\textwidth]{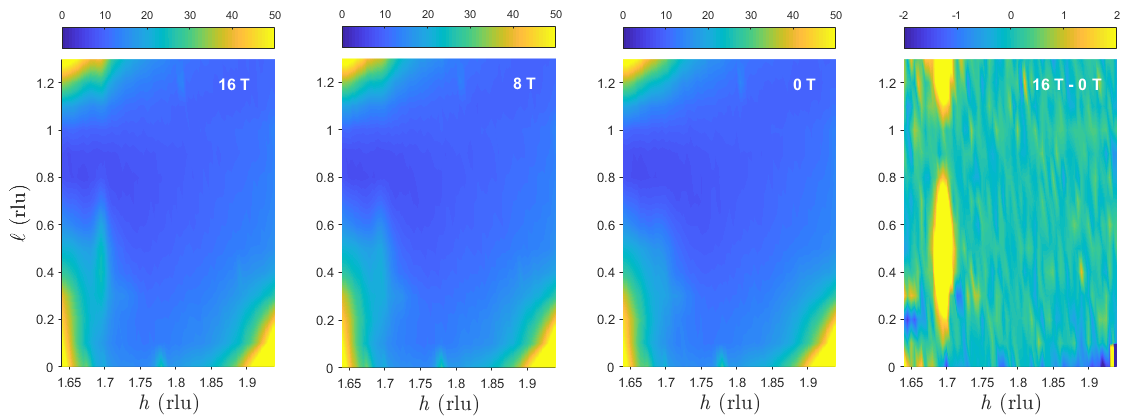}
	\caption{Map of the scattering at ($H$ 0 $L$) at different magnetic fields.  The panel to the right shows the difference between the map at 16 T and 0 T.}
	\label{fig:HLmaps}
\end{figure*}

We focus on the data in ranges close to the $q_{CDW}$ and $q_{PDW}$ positions in Figure \ref{fig:linecuts_all}.  This shows linecuts through the CDW and PDW positions in the $K$ and $H$ directions at $L$ = 0.5 and $L$ = 1 at fields from 0 T to 16 T, focussing on the (2-$q$) positions.  The data have been fitted using a Gaussian function on a sloping background.  Where there is no obvious peak, the centre has been set to the expected CDW or PDW value, and the width has been fitted with an upper limit about 20\% larger than that observed in the cases where the peaks could be fitted.  As expected, for the ($H$ 0 1) case, no high field peak is seen.

\begin{figure}[h]
	\centering
	\includegraphics[width=0.49\textwidth]{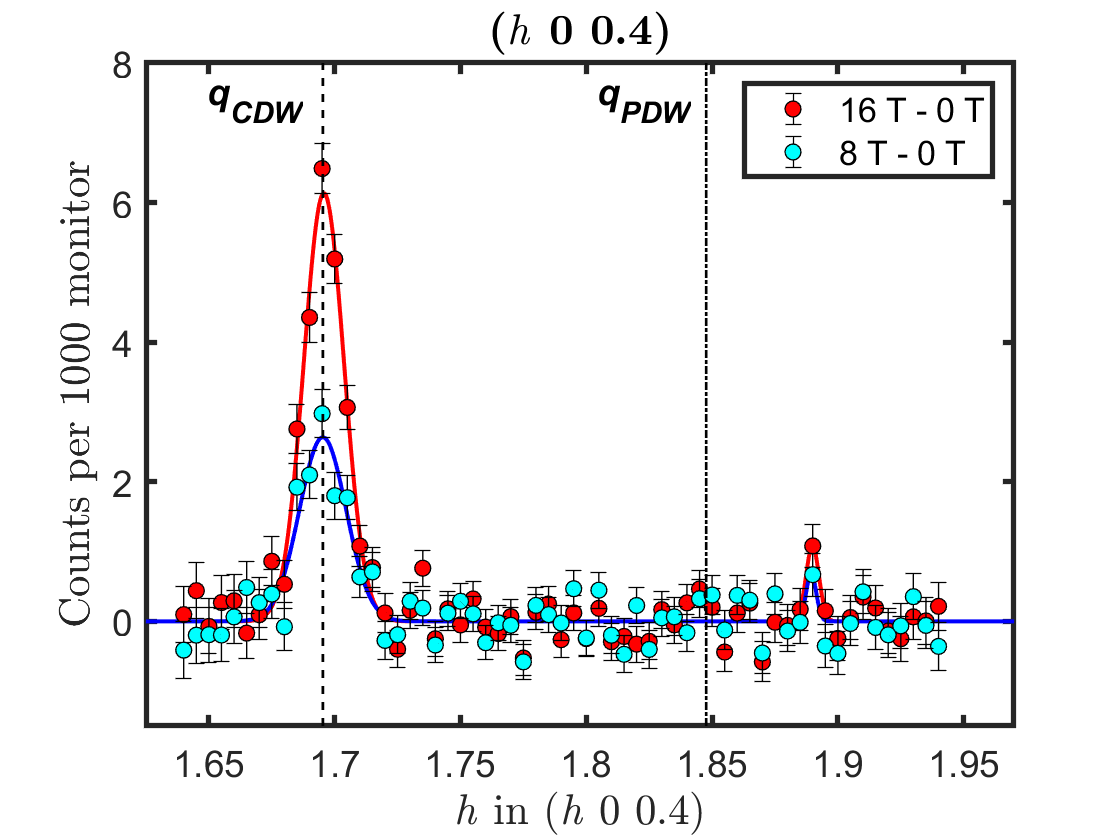}
	\caption{The only non-CDW feature that we observed with a field dependence was a small spike at (1.89 0 0.4). In both $H$ and $L$ cuts this feature consists of one anomalous point.  The relevant $H$ cuts are shown above, with the 0 T data subtracted for clarity.  A monitor of 1000 corresponds to a counting time of 1s.  The dashed line denotes $q_{CDW}$ and the dash-dotted line denotes $q_{PDW} = \frac{1}{2} q_{CDW}$.  The fits are to two Gaussians on a zero background. The widths have been constrained to be greater than the experimental resolution.}
	\label{fig:HLspot}
\end{figure}

To illustrate that a signal is not appearing at a non-rational $L$ value, Figure \ref{fig:HLmaps} presents maps in ($H 0 L$) space, measured at 0 T, 8 T and 16 T, along with the difference between 16 T and 0 T.  Here, we see the CDW signal clearly, especially after subtraction, and the background from the primary structural Bragg peaks and the scattering from the Cu-O chains cancels out nicely.  Close inspection of the data highlight a couple of additional high points in the subtraction.  These are all revealed to be single points high, as illustrated in Figure \ref{fig:HLspot} for the spot appearing at (1.89 0 0.4).  We speculate that this is scattering from a micro-crystallite in the sample.  Although our sample is detwinned, the detwinning is not perfect (99\% detwinned).\cite{Chang2012_CBHC}  The scattering here may be from a small misaligned twin.

\section{Discussion}

Our experiment shows no evidence for a pair density wave associated with the charge density wave seen in YBCO.  This is consistent with observations by Vinograd \textit{et al.} on YBCO under uniaxial pressure. \cite{Vinograd}  This could be because there is no pair density wave in this material, but could also arise for several other reasons.  As discussed above, the X-rays are looking at atomic positions induced by charge density distributions.  This coupling may be too weak in YBCO.  The form factor effect may mean that we are too far away in momentum space to see the effect.

The level of coupling between the $d$-wave "standard" superconducting order parameter and the pair density wave order parameter may be too weak in YBCO.  Our results impose limits on the $\Delta_0$ that are quite tight.

In addition, the CDW observed could be an independent order parameter and any PDW/PDW-induced CDW is elsewhere in reciprocal space.

\end{document}